\documentclass[aps,pra,twocolumn,showpacs]{revtex4-1}

\newcommand{\Jeff}{J_{\mbox{\footnotesize eff}}}
\newcommand{\stat}{\sigma_{\mbox{\footnotesize stat}}}
\newcommand{\erec}{E_{\mbox{\footnotesize rec}}}

\usepackage{graphicx}

\begin{document}

\title{Expansion of matter waves in static
and driven periodic potentials}

\author{C.E.~Creffield, F.~Sols}
\affiliation{Departamento de F\'isica de Materiales, Universidad
Complutense de Madrid, E-28040, Madrid, Spain}
\author{D. Ciampini$^{1,2}$, O.~Morsch$^{2}$, and E.~Arimondo$^{1,2}$}
\affiliation{$^{1}$CNISM-Pisa, Dipartimento di Fisica, Universit\`{a} di
Pisa, Largo Pontecorvo 3, 56127 Pisa, Italy}
\affiliation{$^{2}$INO-CNR, Dipartimento di Fisica, Universit\`{a} di
Pisa, Largo Pontecorvo 3, 56127 Pisa, Italy}

\date{\today}

\pacs{03.75.Lm, 03.65.Xp}

\begin{abstract}
We study the non-equilibrium dynamics of cold atoms held in
an optical lattice subjected to a periodic driving potential.
The expansion of an initially confined atom cloud
occurs in two phases: an initial quadratic expansion followed
by a ballistic behaviour at long times. Accounting for this gives a
good description of recent experimental results, and provides
a robust method to extract the effective intersite tunneling from
time-of-flight measurements.
\end{abstract}

\maketitle

\section{Introduction}
Experimental progress in confining ultracold atoms in optical lattices
has undergone spectacular progress in recent years.
Optical lattice potentials are extremely clean and controllable,
and the excellent coherence properties of atomic condensates hold out
the prospect of controlling their dynamics using quantum coherent
methods. This is both interesting from the point of view of
fundamental physics, and has many potential applications to quantum
information processing, where it is vital that the coherence of
the system is preserved during its time evolution. One such scheme
is to use a time-periodic driving potential to
induce the effect termed ``dynamical localization'' \cite{kenkre} or
``coherent destruction of tunneling'' \cite{hanggi}.
This is a quantum interference effect in which a particle
acquires a phase from its interaction with the driving potential,
which leads to a renormalization of the single-particle tunneling probability.
For specific values of the driving parameters the effective
tunneling probability can be highly suppressed, providing a sensitive
means of coherently controlling the localization of the
atoms \cite{creffield_prl}. This renormalization
has recently been directly observed in cold atom experiments
\cite{pisa_flat,pisa_tilt,oberthaler}.

A convenient way of measuring the effective tunneling is
to observe the rate of expansion of a condensate once a
harmonic potential trapping the atoms along the direction of the
optical lattice has been switched off \cite{pisa_flat, pisa_tilt}.
Although the results of these experiments agreed well with the theoretically
expected scaling of the renormalized tunneling probability with the Bessel
function of the driving strength, for particular
initial conditions the scaling seemed to be {\em quadratic} rather
than linear in the Bessel function.
A number of explanations of this phenomenon have since been put forward,
including a possible crossover from coherent to sequential tunneling
\cite{pisa_tilt,kolovsky} induced by phase scrambling arising
from dynamical instabilities, a time-averaging effect produced
by finite time-resolution of the measurement \cite{prl_2008},
and driving-induced atom-pairing \cite{weiss}.
A recent theoretical stability analysis \cite{creff_pra} has shown, however,
that phase scrambling is unlikely to occur for the experimental parameters
of Refs. \cite{pisa_flat, pisa_tilt}, and the pairing mechanism would require
rather stronger interactions than were present in those experiments.
In this paper we propose an alternative explanation for the
observed quadratic scaling of the renormalized tunneling probability based on
the exact form of the expansion of the condensate, which is linear in the
long time limit but quadratic for short times \cite{textbook}.
Using this complete time dependence in order to extract the tunneling
probability from the experimental data gives quantitatively accurate
agreement with theory, with no adjustable parameters.

\section{Model and Analysis}
The Bose-Hubbard model is described by the Hamiltonian
\begin{eqnarray}
H_{\mbox{\footnotesize BH}} = -J \sum_{\langle i,j \rangle}
\left[ a_i^{\dagger} a_j + \mbox{H.c.} \right]
&+& \frac{U}{2} \sum_j n_j (n_j - 1) \nonumber \\
 &+& \sum_j V(r_j) n_j ,
\label{bose}
\end{eqnarray}
where $a_j / a_j^{\dagger}$ are annihilation / creation operators
for a boson on lattice site $j$, $J$ (taken to be positive) describes
the hopping amplitude between nearest neighbor sites $\langle i, j \rangle$,
and $U$ is the repulsive energy between two bosons occupying the
same site. The operator $n_j = a_j^{\dagger} a_j$ is the standard
number operator, and $V(r)$ is the external trap potential, which is
usually considered to be parabolic,
$V(r) =  m \omega_T^2 r^2 / 2$, where
$\omega_T$ is the trap frequency.
Although simple in appearance, the Bose-Hubbard model can provide an
excellent description \cite{jaksch} of
ultracold atoms held in optical lattice potentials.

Adding a static and a sinusoidally varying force to the system leads
to the general time-dependent potential
\begin{equation}
H(t) = H_{\mbox{\footnotesize BH}} +
\sum_j \ n_j \ j \ \left( \Delta + K \cos \omega t \right) ,
\label{potential}
\end{equation}
where $\Delta$ is the static tilt applied to the lattice,
and $K$ and $\omega$ are the amplitude and frequency, respectively,
of the oscillating component. Experimentally, the two forces are
introduced into the rest frame of the optical lattice by applying
appropriate frequency differences to the acousto-optic modulators
creating the lattice beams, as described in detail in \cite{pisa_flat}.
A time-periodic system of this type can be analyzed using
Floquet theory, revealing \cite{holthaus} that
the effect of the driving can be described by the static
Hamiltonian (\ref{bose}) with a {\em renormalized tunneling}
$\Jeff$. For an untilted lattice ($\Delta = 0$)
this renormalization is given by the zeroth-order Bessel
function $\Jeff = J {\cal J}_0(K_0)$,
where for convenience we define $K_0  \equiv K/\hbar \omega$.
Thus at the values $K_0 = 2.404, \ 5.52, \dots$,
at which the Bessel function vanishes,
the effective tunneling is suppressed.
This effect thus provides a means to coherently
control the dynamics of trapped atoms, without
altering any of the parameters of the optical lattice.

When the trap potential along the lattice direction
is removed the atom cloud will expand in time, at a rate determined
by $\Jeff$. To quantify this process we calculate the spread of the
wavefunction
\begin{equation}
\sigma(t) = \sqrt{ \langle x^2 \rangle - \langle x \rangle^2},
\label{spread}
\end{equation}
aligning the optical lattice with the $x$-axis.
We begin by setting the Hubbard interaction, $U$,
to zero. In the continuum approximation,
valid when the kinetic energy of the
condensate is much less than the width of the
first Bloch band, the ground state of a parabolic
trap is simply given by a Gaussian, $\psi(x) = N \exp[ -x^2/(2 a^2)]$,
where $N$ is the normalization such
that $\int_{-\infty}^{\infty} | \psi(x) |^2 dx = 1$, and
$a = \sqrt{\hbar / (m \omega_T)}$ is the harmonic trap-length.
If the trap potential is now removed, this initial state will
expand while {\em remaining} Gaussian.
This expansion can be calculated analytically
\cite{textbook} yielding the result
\begin{equation}
\sigma(t) = \sigma_0 \ \sqrt{1 +
4 \left(J t/ \hbar \right)^2 \left( d_L / a \right)^4} ,
\label{expand}
\end{equation}
where $d_L$ is the spacing of the optical lattice,
and $\sigma_0 = a/\sqrt{2}$.
Similar, but more complicated, expressions were obtained by
Korsch \cite{korsch} using a lattice representation instead
of the continuum approximation, which coincide with this result.
The expansion clearly occurs in two different phases,
separated by a crossover time
$t_c = \left( \hbar / 2 |J| \right) \left( a / d_L \right)^2$.
For long expansion times, $t \gg  t_c$,
the wavefunction spreads linearly with time,
$\sigma(t) \propto |J| t$, reproducing the expected ballistic
expansion of a released wavepacket. For short times,
however, $t < t_c$, the expansion is instead {\em quadratic},
$\sigma(t) \propto  J^2 t^2$.
It is important to note
that the expansion depends on both the magnitude of the
tunneling, $J$, as well as on the size of the initial
wavepacket $a$. In particular,
a tightly-confined wavepacket will have a higher
spread in momentum, and so will enter the
regime of linear expansion more quickly.
The extreme case where only a {\em single} lattice site
is filled was considered in Ref. \onlinecite{kenkre},
where it was found that the expansion was always linear with time.
This result is exactly reproduced by Eq. (\ref{expand})
by appropriately taking the limit $a \rightarrow 0$.

\section{Results}

\begin{center}
\begin{figure}
\includegraphics[width=0.4\textwidth,clip=true]{fig1a}
\includegraphics[width=0.4\textwidth,clip=true]{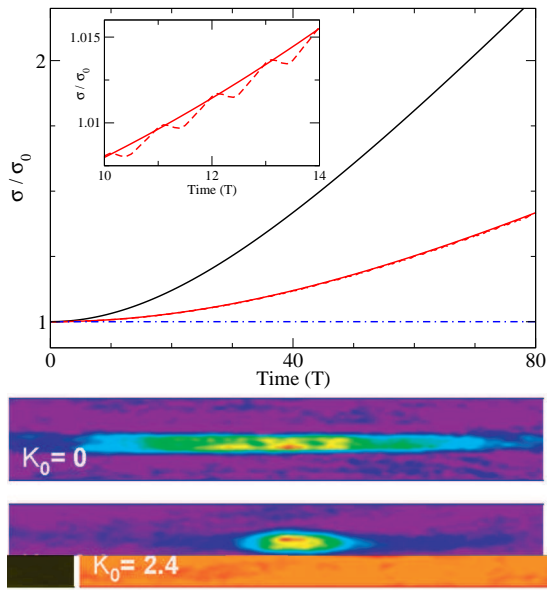}
\caption{(Color online)
Above: Expansion of an initial Gaussian wavepacket in
a flat lattice potential, obtained by the
numerical propagation under Hamiltonian (\ref{potential}).
When no driving potential
is applied ($K = 0$) $\sigma(t)$ increases following
Eq. (\ref{expand}). For $J=0.1\erec$ (black solid line) the expansion
is clearly quadratic initially, and becomes linear at long
times. Setting $J=0.05 \erec$ (red solid line) reduces the
expansion rate as expected.
By applying a periodic driving potential the tunneling can be renormalized
to an effective value $\Jeff$. Tuning $K_0 = 1.22$
reduces $\Jeff$ so that the expansion of the
condensate (dashed red line) reproduces the $J=0.05 \erec$ result.
Setting $K_0 = 2.404$ -- the first zero of the Bessel function --
produces coherent destruction of tunneling (CDT), and the condensate
no longer expands with time (blue dash-dotted line).
{\em Inset - } Detail of the periodically-driven result
($K_0 = 1.22$). The driven expansion {\em on average} reproduces
the $J=0.05 \erec$ result, but
shows small oscillations with the same frequency of the
driving. The amplitude of these oscillations decreases
with increasing driving frequency.
Below: Experimental comparison of the free expansion of a condensate
($K_0 = 0$) with a condensate experiencing CDT ($K_0 = 2.4$). As
predicted, the expansion of the second condensate is strongly
suppressed.}
\label{expansion}
\end{figure}
\end{center}

{\em Undriven lattice --}
We first consider the case of a static lattice,
with $\Delta, K_0 = 0$.
In Fig. \ref{expansion} we show the time-dependence of the
expansion of an initial Gaussian wavepacket,
numerically evolved in time under the
time-dependent Hamiltonian (\ref{potential}).
Using $J=0.1 \erec$, where the recoil energy
$\erec = \hbar^2 \pi^2/ 2 m d_L^2$,
we see that the analytic expression (\ref{expand})
accords exactly with the numerical result,
indicating the validity of the continuum approximation.
The transition from the initial quadratic expansion to the
ballistic regime is clearly visible.
Halving the value of $J$ used produces the expected
result of reducing the rate of expansion, and moving
the crossover from the quadratic to the ballistic regime
to a later time.

{\em Untilted driven lattice -- }
We now consider the effect of including the time-dependent
driving potential $V(t) = K \cos \omega t$.
We choose a high driving frequency,
$\hbar \omega = 4 J$, and tune the amplitude of the driving
so that $\Jeff = 0.5 J$. In accordance with
the predictions of the Floquet analysis, we see
that the expansion of the wavepacket in the driven system with
a bare tunneling of $J=0.1 \erec$
closely follows the result for $J=0.05 \erec$, indicating that
the driving field indeed renormalizes the tunneling as expected.
Looking in detail at the expansion (inset of Fig. \ref{expansion})
we see that although on average the expansion of the driven condensate
closely matches that of the static case with $J=0.05 \erec$, the
driven result contains small amplitude oscillations with the
same frequency as the driving. These oscillations arise from the
intrinsic time dependence of the Floquet states themselves.
Their amplitude reduces as the driving frequency becomes
larger, indicating that the approximation of modelling
the driven system with a renormalized static Hamiltonian
becomes increasingly good.
Finally we also show in Fig. \ref{expansion} the most dramatic effect
of the renormalization of tunneling.
Since $\Jeff = J {\cal J}_0(K_0)$, tuning $K_0$
to a zero of the Bessel function should result in the
complete suppression of tunneling (neglecting next-nearest neighbor
tunneling \cite{pisa_oldenburg}). We indeed see that setting
$K_0 = 2.404$ results in the condensate not
expanding with time, due to the vanishing of $\Jeff$.
Similarly to the $K_0 = 1.22$ case, this curve
again displays small oscillations, which become larger
at low values of $\omega$. This low frequency behaviour
would correspond to the ``dynamical localization''
regime \cite{kenkre}, where the wavepacket periodically
returns to its initial state at stroboscopic times $t = n T = n 2\pi/\omega$,
but between these times can exhibit large excursions.

\begin{center}
\begin{figure}
\includegraphics[width=0.4\textwidth,clip=true]{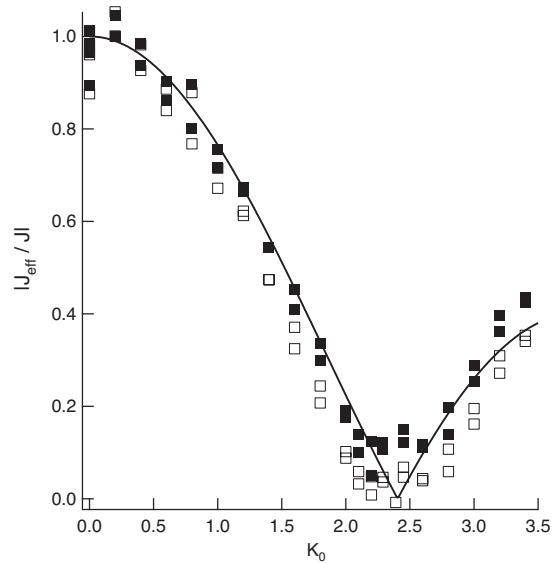}
\caption{Dynamical suppression of tunneling in a periodically
driven lattice. The effective tunneling parameter was calculated
assuming a simple linear expansion (open symbols) and by taking into
account the exact expansion dynamics (filled symbols). In the latter
case the agreement with the theoretically expected scaling (solid line)
is clearly better.
The experimental parameters were $J/h= 240\,\mathrm{Hz}$,
$\omega/2\pi=4\,\mathrm{kHz}$ and the expansion time $150\,\mathrm{ms}$.}
\label{J0}
\end{figure}
\end{center}

In Ref. \onlinecite{pisa_flat} the effective tunneling was deduced
by measuring the expansion rate of the condensate at a fixed
time, and assuming this rate was directly proportional to $\Jeff$.
Accordingly the ratio between the tunneling parameters in the static
and the driven lattice was calculated as
$|\Jeff/J|=(\sigma(t)-\sigma_0)/(\stat-\sigma_0)$,
where $\stat$ is the size of the condensate after expansion
in the static lattice. For the experimental parameters
($d_L = 426$ nm and $J / h = 270$ Hz),
we calculate a crossover time of $t_c \simeq 9.7$ ms
for a weakly-interacting condensate released from
a 20 Hz harmonic trap. As the experiment employed an expansion
time of 100 ms, the results can thus be expected to
be reliable only as long as $| \Jeff / J| \geq 0.1$. In order to get better
agreement with theory we now use Eq. (\ref{expand}) containing the full
expansion dynamics, giving
\begin{equation}
|\Jeff/J|=\sqrt{\frac{\sigma(t)^2-\sigma_0^2}{\sigma_{stat}^2-\sigma_0^2}}.
\label{full}
\end{equation}
Figure \ref{J0} shows that, as expected, using Eq. (\ref{full}) to
calculate the renormalized tunneling gives better agreement with theory.

It is interesting to note that although $\Jeff$ is strongly
suppressed near $K_0 = 2.4$, it does not actually
reach zero when the Bessel function vanishes. This is due to
the effect of higher-order hopping terms present in the
system's dynamics. Although they too are renormalized by the driving
potential, for sinusoidal driving they will not vanish at the same driving parameters
as for the nearest neighbor hopping. The residual
value of $\Jeff$, visible in Fig. \ref{J0}, is in reasonable agreement with the value of the next-to-nearest neighbor hopping (around 5$\%$)
calculated for a similar system in Ref.\cite{pisa_oldenburg}.

\begin{center}
\begin{figure}
\includegraphics[width=0.40\textwidth,clip=true]{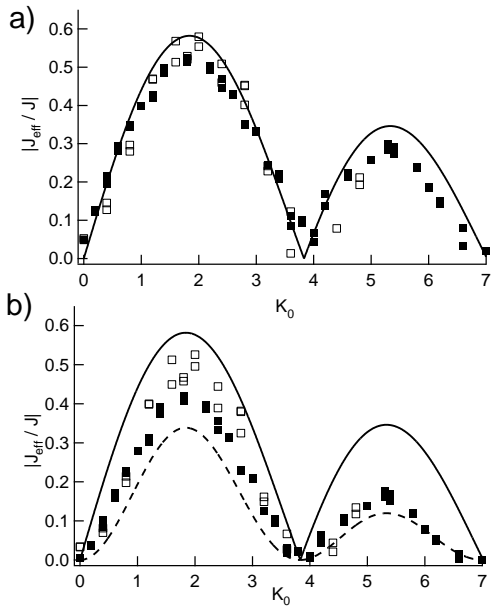}
\caption{a) Effective tunneling for resonant driving in a tilted lattice
(photon-assisted tunneling).
The effective tunneling parameters were calculated for
two different initial condensate sizes (around $15\,\mathrm{\mu m}$
(open symbols) and around
$17\,\mathrm{\mu m}$ (filled symbols)) using Eq. (\ref{expand}).
For comparison, b) shows the same experimental data with the
renormalized tunneling parameter calculated assuming linear expansion.
One clearly sees that in this case the experimental data interpolate
between a linear Bessel scaling (solid line) and a quadratic scaling (dashed line).}
\label{fig3}
\end{figure}
\end{center}

{\em Tilted lattice, resonant driving --}
Applying Eq. (\ref{full}) to the experimental data on photon-assisted
tunneling \cite{pisa_tilt} leads to an even more striking improvement
in the agreement between theory and experiment. In those experiments a
tilt was applied to the lattice through a constant acceleration,
leading to a suppression of tunneling by Wannier-Stark localization.
Periodic driving of the lattice at a frequency $\omega$ matching the energy
offset between two adjacent lattice wells then led to partial restoration
of the tunneling probability, with the effective tunneling probability
given by $|\Jeff/J|= {\cal J}_1(K_0)$, i.e. one expects a scaling
with the first-order Bessel function. As shown in the inset of Fig. \ref{fig3},
assuming linear expansion in order to extract $\Jeff/J$ led to a scaling
that interpolated between a linear and a quadratic dependence
on ${\cal J}_1(K_0)$, depending on the initial size of the condensate
(which in the experiment was varied through the nonlinearity by changing
the atom number). If the full expansion dynamics is taken into account
through Eq. (\ref{full}), however, both data sets give the same dependence
on $K_0$ which is very close to the theoretical prediction.

\section{Conclusions}
We have shown that in order to extract the effective tunneling from
expansion measurements it is important to account
for the detailed time-dependence of the condensate expansion.
When the effective tunneling is small, or the initial width
of the condensate is large, the crossover
to ballistic expansion will not be reached until very long times.
Measurements made at earlier times will thus underestimate the
effective tunneling rate, which gives a quantitatively accurate
interpretation of the ``squared Bessel function''
behaviour noted in Refs. \onlinecite{pisa_flat, pisa_tilt}.
Although we have not included the effects of interactions,
this is a reasonable approximation for the systems studied
in Refs. \onlinecite{pisa_flat,pisa_tilt} where the interactions
were fairly small ($U / h \simeq 10$ Hz), and their effect
rapidly became negligible as the condensate expanded and became
more dilute. Using the correct expansion formula, Eq. (\ref{expand}),
not only provides an accurate means of deducing the
value of the effective tunneling, but is essential to
study subtle effects, such as the transition to diffusive
tunneling and the influence of higher-order tunneling terms,
which would otherwise be masked by this behaviour
when the effective tunneling is small.

\acknowledgments
This research was supported by the Acc\'ion Integrada /
Azioni Integrate scheme (Spain-Italy).
The authors also acknowledge support from the Spanish MICINN through
Grant No. FIS-2007-65723 and the Ram\'on y Cajal program (CEC).
We thank H. Lignier,
C. Sias, Y. Singh, and A. Zenesini for assistance with the experiments.


\begin{thebibliography}{99}

\bibitem{kenkre}
{D.H.~Dunlap and V.M.~Kenkre, Phys. Rev. B {\bf 34}, 3625 (1986).}

\bibitem{hanggi}
{F.~Grossmann, T.~Dittrich, P.~Jung, and P.~H\"anggi, Phys. Rev. Lett.
{\bf 67}, 516 (1991).}

\bibitem{creffield_prl}
{C.E.~Creffield, Phys. Rev. Lett. {\bf 99}, 110501 (2007).}

\bibitem{pisa_flat}
{H.~Lignier, C.~Sias, D.~Ciampini, Y.~Singh, A.~Zenesini,
O.~Morsch, and E.~Arimondo, Phys. Rev. Lett. {\bf 99}, 220403 (2007).}

\bibitem{pisa_tilt}
{C.~Sias, H.~Lignier, Y.P.~Singh, A.~Zenesini, D.~Ciampini,
O.~Morsch, and E.~Arimondo, Phys. Rev. Lett. {\bf 100}, 040404 (2008).}

\bibitem{oberthaler}
{E.~Kierig, U.~Schnorrberger, A.~Schietinger, J.~Tomkovic,
and M.K.~Oberthaler, Phys. Rev. Lett. {\bf 100}, 190405 (2008).}

\bibitem{kolovsky}
{A.R. Kolovsky and H.J.~Korsch, arXiv:0912.2587.}

\bibitem{prl_2008}
{C. E. Creffield and F. Sols, Phys. Rev. Lett. {\bf 100}, 250402 (2008).}

\bibitem{weiss}
{C.~Weiss and H.-P.~Breuer, Phys. Rev. A {\bf 79}, 023608 (2009).}

\bibitem{creff_pra}
{C.E.~Creffield, Phys. Rev. A {\bf 79}, 063612 (2009).}

\bibitem{textbook}
{A.~Galindo and P.~Pascual, {\em Quantum Mechanics I (Theoretical
and Mathematical Physics)}, Ch. 3 (Springer-Verlag, 1990).}

\bibitem{jaksch}
{D.~Jaksch, C.~Bruder, J.I.~Cirac, C.W.~Gardiner,
and P.~Zoller, Phys. Rev. Lett. {\bf 81}, 3108 (1998).}

\bibitem{holthaus}
{M.~Holthaus, Phys. Rev. Lett. {\bf 69}, 351 (1992).}

\bibitem{korsch}
{H.J.~Korsch and S.~Mossmann, Phys. Lett. A {\bf 317}, 54 (2003);
A.~Klumpp, D.~Witthaut, and H.J.~Korsch J. Phys. A: Math. Theor.
{\bf 40}, 2299 (2007).}

\bibitem{pisa_oldenburg}
{A. Eckardt {\em et al.}, Phys. Rev. A {\bf 79}, 013611 (2009).}

\end{thebibliography}
\end{document}